\documentclass [12pt]{article}
\def\maj#1{\ifmmode\mbox{\usefont{U}{msb}{m}{n}#1}\else{\usefont{U}{msb}{m}{n}#1}\fi}
\def\v#1{\mathbf{#1}}
\usepackage{graphics,epsfig,graphicx}
\usepackage{color}

\begin{document}

\title{\textbf{Polariton-polariton scattering :\\ exact results through
a novel approach}}
\author{\small{M. Combescot$^1$, M.A. Dupertuis$^{1,2}$ and O.
Betbeder-Matibet$^1$}\\
$^1$\small{\it{Institut des NanoSciences de Paris (INSP),}} \\
\small{\it{Universit\'e Pierre et Marie Curie-Paris 6, Universit\'{e}
Denis Diderot-Paris 7, CNRS, UMR 7588,}} \\ \small{\it{Campus Boucicaut,
140 rue de Lourmel, 75015 Paris}}\\
  $^2$ \small{\it{Laboratoire d'Opto\'electronique Quantique,}}\\
\small{\it{
Ecole Polytechnique F\'ed\'erale de Lausanne (EPFL),}}\\
\small{\it{Station 3,
CH-1015 Lausanne, Switzerland}}}
\date{}
\maketitle

\begin{abstract}
We present a fully microscopic approach to the transition rate of two
exciton-photon polaritons. The non-trivial consequences of the
polariton composite nature --- here treated exactly through a
development of our composite-exciton many-body theory --- lead to results
noticeably different from the ones of the conventional approaches in
which polaritons are mapped into elementary bosons. Our work reveals an
appealing fundamental scattering which corresponds to a photon-assisted
exchange --- in the absence of Coulomb process. This scattering being
dominant  when one of the scattered polaritons has a strong photon
character, it should be directly accessible to experiment. In the case of
microcavity polaritons, it produces a
significant enhancement of the polariton transition rate when compared to
the one coming from  Coulomb interaction. This paper also contains the
crucial tools to securely tackle the many-body physics of polaritons, in
particular towards its possible BEC.
\end{abstract}

\vspace{0.5 cm}


71.35.-y Excitons and related phenomena 

71.35.Lk Collective effects (Bose effects, phase space filling, and
excitonic phase transitions)

\newpage

Almost 50 years ago, J.J. Hopfield [1] has shown that photons can be
dressed by semiconductor excitations to form a mixed state, the
polariton. As this mixed state is the exact eigenstate of the coupled
photon-semiconductor Hamiltonian, semiconductor should not absorb any
photon. Actually, photon
absorption is observed when the exciton broadening is large compared to
the exciton-photon coupling. The absorption
then follows the Fermi golden rule in
spite of the fact that the transition is not towards a
continuum, but a discrete state, namely, the exciton with a
center-of-mass momentum equal to the photon momentum. This can be
physically understood by saying that the exciton broadening acts
as a continuum around the exciton discrete level [2].

In the most basic approach to polaritons [3,4], the excitons are
considered as non-interacting elementary bosons: The polariton effect
then is a bare one-photon one-exciton effect, independent from the
laser intensity. This has to be contrasted with the exciton optical Stark
effect discovered 25 years ago by D. Hulin and coworkers [5,6], in which
the observed exciton line shift, proportional to the laser intensity, only
comes from excitons differing from elementary bosons. This
apparent contradiction in problems quite similar, namely,
photons interacting with a semiconductor, has been tackled by one of us
[7] : Through the derivation of the polariton effect and the exciton
optical Stark effect within the same framework, it is possible to show
that the polariton picture with excitons as non-interacting
elementary bosons, is valid at low laser intensity, while the composite
nature of the excitons becomes crucial when the laser intensity gets
large. 

We have recently constructed a many-body theory for excitons in which the
composite nature of the particles is treated exactly. It has, up to now,
been developed through more than 30 publications. This leads us to
consider that the reader now has some knowledge of this many-body theory,
at least through the short review paper [8] or, for more details, through
the appendices of ref.\ [9]. The most important result of this theory is
the fact that excitons interact not only through Coulomb scatterings for
interactions between their carriers, but also through ``Pauli
scatterings'' for
\emph{carrier exchanges in the absence of Coulomb process}. It turns out
that these dimensionless pure carrier exchanges --- which are by
construction missed in effective Hamiltonians for boson excitons, due to
a bare dimensional argument, --- dominate
\emph{all} semiconductor optical nonlinearities. Among these nonlinear
effects, we can cite the exciton optical Stark shift [6] extensively
studied in the 80's. We can also cite the Faraday rotation in photoexcited
semiconductors that we have recently suggested [10]. More generally,
through a novel approach to the nonlinear susceptibility [11], we have
proved that the large detuning behavior of
$\chi^{(3)}$ is \emph{entirely} controlled by these pure carrier exchanges
between composite excitons. Processes in which Coulomb interactions
take place only enter as detuning corrections.

An ``ideal'' polariton made of non-interacting boson excitons and
(non-interacting boson) photons, would behave as a non-interacting boson.
In order to explain the observed scatterings between polaritons, it is
necessary to take into account the fact that excitons do interact. To
possibly treat their interactions through available many-body procedures
--- valid for elementary quantum particles only [12,13], ---
excitons are commonly considered as elementary bosons, their composite
nature being hidden through the Coulomb exchange term of the
exciton-exciton scattering in the exciton effective Hamiltonian generated
by the ``bosonization'' procedure. In previous works, we have shown that
not only the effective exciton-exciton scattering up to now used [14]
should have been rejected long ago because it induces an unphysical non
hermiticity in the effective exciton Hamiltonian [15], but, worse, there
is no way to properly treat the interactions between excitons through an
effective potential between elementary bosons, whatever the
exciton-exciton scatterings are [9,16]. All our works on
exciton many-body effects end with the same conclusion : It is not
possible to forget the composite nature of the excitons by reducing them
to elementary bosons, whatever the bosonization procedure is.

The present work on polariton-polariton scatterings once more supports
this conclusion. Our composite-exciton many-body theory reveals the
existence of an appealing ``photon-assisted exchange'' channel in the
scattering of two polaritons which directly comes from the exciton
composite nature (see fig.1(a)) : In this channel, the excitonic parts
$i_1$ and
$i_2$ of the  polaritons
$P_1$ and $P_2$ exchange their carriers without any Coulomb process. In a
second step, one of the resulting excitons, let us say
$i'_1$, transforms into the photon part $n'_1$ of the
polariton $P'_1$ through the
vacuum Rabi coupling, so that this photon-assisted exchange scattering
also is an energy-like quantity in spite of the absence of Coulomb
process. When compared to the two other channels of the standard approach
to polariton-polariton scattering,  associated to direct and exchange
Coulomb interaction between the excitonic parts of the polaritons (see
figs.1(b,c))), this  photon-assisted exchange scattering is obviously
going to be dominant when one of the four polaritons has a strong
photon character, the photon-polariton overlap being then
larger than the exciton-polariton overlap. This makes it directly
accessible to experiments.

\noindent\textbf{Microscopic formalism}

The Hamiltonian of a semiconductor coupled to a
photon  field splits as
$H=H_{\mathrm{ph}}+H_\mathrm{sc}+H_{\mathrm{ph-sc}}$.
The photon part reads $H_\mathrm{ph}=\sum_n\omega_n 
a_n^\dag a_n$ where $a_n^\dag$ creates a photon in a mode 
$n$. The semiconductor part,
$H_\mathrm{sc}=H_e+H_h+V_{ee}+V_{hh}+V_{eh}$, contains 
kinetic and Coulomb contributions for \emph{free} carriers. Its
one-electron-hole-pair eigenstates,
$(H_\mathrm{sc}-E_i)B_i^\dag|v\rangle=0$, are the excitons, their
bound and extended states forming a complete basis for one-pair
states.
For photons close to the exciton
resonance, the photon-semiconductor coupling,
$H_\mathrm{ph-sc}=W+W^\dag$, can be reduced to its resonant terms,
$W=\sum_{n,i}\Omega_{n i}\,\alpha_n^\dag B_i$,
where $\Omega_{n i}$ is the vacuum Rabi coupling between the
$n^\mathrm{th}$ photon mode and the $i^\mathrm{th}$ exciton: $W$ creates a
photon while destroying an exciton.

Polaritons $C_P^\dag|v\rangle$ are the $H$ eigenstates in
the subspace made of one photon coupled to one  exciton, 
$(H-\mathcal{E}_P)C_P^\dag|v\rangle=0$. They thus form a  complete
orthogonal basis for this subspace. Consequently, we can write photons 
and excitons in terms of polaritons as
\begin{equation}
a_n^\dag=\sum_PC_P^\dag\,\alpha_{Pn}\ ,\ \ \ \
B_i^\dag=\sum_PC_P^\dag\,\beta_{Pi}\ ,
\end{equation}
while polaritons read in terms of photons and excitons as
\begin{equation}
C_P^\dag=\sum_n
a_n^\dag\,\alpha_{nP}+\sum_iB_i^\dag\,\beta_{iP}\ .
\end{equation}
The prefactors in these expansions are nothing but the Hopfield
coefficients, \emph{i.e.}, the photon-polariton overlap
$\alpha_{nP}=\langle v|a_nC_P^\dag|v\rangle=\alpha_{Pn}^\ast$ and the
exciton-polariton overlap
$\beta_{iP}=\langle v|B_iC_P^\dag|v\rangle=\beta_{Pi}^\ast$. They can be
made large or small depending on the polariton character.

\noindent\textbf{Elementary scatterings between polaritons}

Due to their excitonic components, the polaritons are not
exact bosons. This is readily seen from
$[C_{P'},C_P^\dag]=\delta_{P',P}-\tilde{D}_{P'P}$, where
$\tilde{D}_{P'P}$ is the ``polariton deviation-from-boson operator''.
Using eqs.\ (1,2), $\tilde{D}_{P'P}$ is
just the exciton deviation-from-boson operator
$D_{i'i}$ of the composite-exciton many-body theory (see eq.\
(1.3) of ref.\ [9]), dressed by photons through the exciton-polariton
overlaps, namely, 
$\tilde{D}_{P'P}=\sum_{i',i}\beta_{P'i'}\,\,D_{i'i}
\,\beta_{iP}$. The Pauli scatterings of two polaritons
$\tilde{\lambda}$, resulting from fermion exchanges as shown in fig.1(d),
appear through 
\begin{equation}
\left[\tilde{D}_{P'_1P_1},C_{P_2}^\dag\right]=\sum_{P'_2}C_{P'_2}^\dag
\left\{\tilde{\lambda}\left(^{P'_2\ P_2}_{P'_1\
P_1}\right)+(P_1\leftrightarrow P_2)\right\}\ .
\end{equation}
These scatterings are just the exciton Pauli scatterings $\lambda$ of the
composite-exciton many-body theory (see eq.\ (1.2) of ref.\
[9]), dressed by photons
\begin{equation} \label{eq:lambdaScattPol}
\tilde{\lambda}\left(^{P'_2\ P_2}_{P'_1\
P_1}\right)=\sum_{i'_1,i'_2,i_1,i_2}\beta_{P'_1i'_1}\beta_{P'_2i'_2}
\,\lambda\left(^{i'_2\ i_2}_{i'_1\ i_1}\right)\,
\beta_{i_2P_2}\beta_{i_1P_1}\ .
\end{equation}

As for
excitons [9], it is not possible to describe the interactions between
polaritons through a potential, due to the composite nature of the
excitonic part of these polaritons. The clean way to overcome this
difficulty is to introduce the
``creation-potential'' of the polariton $P$,
defined as 
\begin{equation}
[H,C_P^\dag]-\mathcal{E}_PC_P^\dag=\tilde{V}_P^\dag-X_P^\dag\ .
\end{equation}
Its first part $\tilde{V}_P^\dag$ barely is the exciton
creation-potential $V_i^\dag$ coming from Coulomb interaction between
excitons (see eq.\ (1.7) of ref.\ [9]), dressed by photons, 
$\tilde{V}_P^\dag=\sum_iV_i^\dag\,\beta_{iP}$.
The second part $X_P^\dag$ is conceptually new. It
comes from the composite nature of the excitons,
through the exciton deviation-from-boson operator $D_{i'i}$
\begin{equation}
X_P^\dag=\sum_{n,i',i}a_{n}^\dag\,\Omega_{ni'}\,D_{i'i}\,\beta_{iP}
\ .
\end{equation}

These two creation-potentials give rise to two physically different
scatterings. The ones associated to $\tilde{V}_P^\dag$,
\begin{equation}
\left[\tilde{V}_{P_1}^\dag,C_{P_2}^\dag\right]=\sum_{P'_1,P'_2}
C_{P'_1}^\dag C_{P'_2}^\dag\,\tilde{\xi}^\mathrm{dir}
\left(^{P'_2\ P_2}_{P'_1\ P_1}\right)\ ,
\end{equation}
shown in Fig.1(b), are na\"{\i}ve. They just correspond to the exciton
direct Coulomb scatterings
$\xi^\mathrm{dir}$ (see eq.\ (1.8) in ref.\ [9]), dressed by
photons as in eq.\ (4),
\begin{equation} \label{eq:directScattPol}
\tilde{\xi}^\mathrm{dir}\left(^{P'_2\ P_2}_{P'_1\
P_1}\right)=\sum_{i'_1,i'_2,i_1,i_2}\beta_{P'_1i'_1}\beta_{P'_2i'_2}
\,\xi^\mathrm{dir}\left(^{i'_2\ i_2}_{i'_1\ i_1}\right)\,
\beta_{i_2P_2}\beta_{i_1P_1}\ .
\end{equation}

The ones associated to the second creation-potential $X_P^\dag$ are more
interesting. They correspond to the photon-assisted exchange  scatterings
dexcribed in the introduction and shown in fig.1(a). Being defined through
\begin{equation}
\left[X_{P_1}^\dag,C_{P_2}^\dag\right]=\sum_{P'_1,P'_2}C_{P'_1}^\dag 
C_{P'_2}
^\dag\left\{\chi\left(^{P'_2\ P_2}_{P'_1\ P_1}\right)+(P_1\leftrightarrow
P_2)\right\}\ ,
\end{equation}
their mathematical expression is
\begin{eqnarray}
\chi\left(^{P'_2\ P_2}_{P'_1\ P_1}\right)=\sum_{
n'_1,i'_1,i'_2,i_1,i_2}\hspace{4cm}\nonumber\\
\beta_{P'_2i'_2}\alpha_{P'_1n'_1}\Omega_{n'_1i'_1}
\lambda\left(^{i'_2\ i_2}
_{i'_1\ i_1}\right)\,
\beta_{i_1P_1}\beta_{i_2P_2}\ .
\end{eqnarray}
They read in terms of the exciton pure exchange Pauli scatterings
$\lambda$, \emph{without any Coulomb process}.

The two energy-like scatterings $\tilde{\xi}^\mathrm{dir}\left(^{P'_2\
P_2}_{P'_1\ P_1}\right)$, $\chi\left(^{P'_2\ P_2}_{P'_1\ P_1}\right)$ and
the dimensionless Pauli scattering $\tilde{\lambda}
\left(^{P'_2\ P_2}_{P'_1\ P_1}\right)$ this procedure generates,
constitute the crucial tools to, in the future, tackle any problem
dealing with the many-body physics of polaritons, with the exciton
composite nature included in an exact way. This is going to be of
importance in view of the claimed recent observation of the polariton
BEC [17,18].

\noindent \textbf{Polariton transition rate}

To get the polariton transition rate, we barely follow the procedure we
have already used to get the exciton transition rate [9,16].
When $H$ does not
split as $H_0+V$, the standard form of the Fermi
golden rule cannot be used. The transition rate from a normalized state
$|\phi_i\rangle$ to a normalized state
$|\phi_f\rangle$, is then obtained through [16]
\begin{equation}
\frac{t}{T_{i\rightarrow f}}=\left|\langle\phi_f|F_t(\hat{H})P_{\perp}H|
\phi_i\rangle\right|^2\ ,
\end{equation}
with $P_{\perp}=1-|\phi_i\rangle\langle\phi_i|$ and
$\hat{H}=H-\langle\phi_i|H|\phi_i\rangle$,
while
$|F_t(E)|^2=2\pi\,t\,\delta_t(E)$ where $\delta_t(E)=(\sin Et/2)/\pi E$
is a peaked function of width $2/t$.

We here consider the transition rate of two identical polaritons $P$
towards two polaritons $(P_1,P_2)$.  To normalize the initial state
$C_P^{\dag 2}|v\rangle$ and the scattered state
$C_{P_1}^\dag C_{P_2}^\dag |v\rangle$, we use
\begin{eqnarray}
\langle v|C_{P'_1}C_{P'_2}C_{P_2}^\dag C_{P_1}^\dag|v\rangle=
\delta_{P'_1,P_1}\,\delta_{P'_2,P_2}-\tilde{\lambda}
\left(^{P'_2\ P_2}_{P'_1\ P_1}\right)\nonumber\\
+\ (P_1\leftrightarrow P_2)\ ,
\end{eqnarray}
which follows from eq.\ (3).

The transition rate from $(P,P)$ to $(P_1,P_2)$, given in eq.\
(11), makes use of $P_{\perp}HC_P^{\dag 2}|v\rangle$. To get it, we note
that, due to eqs.\ (5,7,9), $|S_P\rangle=(H-2\mathcal{E}_P)C_P^{\dag
2}|v\rangle$ reads in terms of the two elementary scatterings of the
polariton many-body theory as
\begin{equation}
|S_P\rangle=
\sum_{P'_1,P'_2}\left\{\tilde{\xi}^\mathrm{dir}\left(^{P'_2\ P}
_{P'_1\ P}\right)-2\chi\left(^{P'_2\ P}_{P'_1\ P}\right)\right\}
C_{P'_1}^\dag
C_{P'_2}^\dag|v\rangle\ .
\end{equation}
As $P_{\perp}C_P^{\dag 2}|v\rangle=0$, this makes
$P_{\perp}HC_P^{\dag 2}|v\rangle$, equal to $P_{\perp}|S_P\rangle$, 
linear in polariton scatterings --- its precise value being $[|S_P\rangle
-|S'_P\rangle]$ with
$|S'_P\rangle=\xi_PC_P^{\dag 2}|v\rangle$, where 
$\xi_P=\langle v|C_P^2|S_P\rangle\langle v|C_P^2C_P^{\dag
2}|v\rangle^{-1}$.  Consequently, to get the transition rate from
$(P,P)$ to
$(P_1,P_2)\neq (P,P)$ at lowest order in the interactions, we just have,
in eq.\ (11), 
to replace $\hat{H}$ by its zero order contribution, i.e., $\hat{H}$ by
$(\mathcal{E}_{P_1}+\mathcal{E}_{P_2}-2\mathcal{E}_P)$ in 
$\langle v|C_{P_2}C_{P_1}F_t(\hat{H})$. We then note that, for a large
sample, the $\tilde{\lambda}$'s, as the exciton Pauli scatterings
$\lambda$, are small compared to 1. This makes the $\tilde{\lambda}$'s in
the normalization factors negligible as well as the contribution coming
from $|S'_P\rangle$. This allows to show that the transition rate from
polaritons
$(P,P)$ to polaritons $(P_1,P_2)$ reduces to [19]
\begin{eqnarray} 
\frac{1}{T_{PP\rightarrow
P_1P_2}}\simeq 4\pi\,\delta_t(\mathcal{E}_{P_1}
+\mathcal{E}_{P_2}-2\mathcal{E}_P)\hspace{3cm}\nonumber\\  \times
\left|\tilde{\xi}^\mathrm{dir}\left(^{P_2\
P} _{P_1\ P}\right)-\tilde{\xi}^\mathrm{in}\left(^{P_2\ P} _{P_1\
P}\right)-\chi\left(^{P_2\ P}_{P_1\ P}\right) -\chi\left(^{P_1\
P}_{P_2\ P}\right)\right|^2\ .
\end{eqnarray}
$\tilde{\xi}^\mathrm{in}$, the diagram of which is shown in
fig.(1c), is the exciton ``in'' Coulomb exchange scattering 
$\xi^\mathrm{in}$ of the composite-exciton many-body theory (see eq.\
(1.10) in ref.\ [9]), dressed by photons as
in eq.\ (8). 

Equation (14) for the polariton transition rate is one of the key results
of the paper. This transition rate conserves
energy at the scale $1/t$, as expected. Its
amplitude contains direct and ``in'' exchange
Coulomb scatterings similar to the ones we found for the exciton
transition rate [9,16]. However, it contains in addition a contribution
from the photon-assisted exchange channel, independent from any Coulomb
process. This physically appealing contribution is directly linked
simultaneously to the exciton composite nature and to the partly photon
nature of the polariton.

\noindent \textbf{State of the art}

In the most na\"{\i}ve approach to polariton-polariton scattering, one
just replaces the excitons in the photon-semiconductor coupling $W$ 
by elementary bosons
$\bar{B}_i$ with $[\bar{B}_i,\bar{B}_j^\dag]=\delta_{i,j}$ and the
semiconductor Hamiltonian by an effective exciton-exciton Hamiltonian. If
the effective scattering $\xi^\mathrm{dir}\left(^{n\
\,j}_{m\ i}\right)-
\xi^\mathrm{out}\left(^{n\ \,j}_{m\ i}\right)$, derived by Haug and
Schmitt-Rink [14], were used, the bracket in the
polariton transition rate (14) would be
$\tilde{\xi}^\mathrm{dir}\left(^{P'_2\ P}_{P'_1\
P}\right)-\tilde{\xi}^\mathrm{out}\left(^{P'_2\ P}_{P'_1\ P}\right)$.
Besides the fact that $\xi^\mathrm{in}\neq \xi ^\mathrm{out}$ (for a
complete discussion, see ref.\ [9]), this
procedure totally misses the photon-assisted exchange scattering
$\chi$ which is dominant when one of the two scattered polaritons has a
strong photon character.

A more elaborate approach, which relies on a truncated Usui's bosonization
procedure, has been proposed by the Quattropani's group [20,21]. Their
Coulomb contribution is now correct  
(the prefactors of eq.\ (18) in ref.\ [20], given in their eqs.\
(19,20), being nothing but $\xi^\mathrm{dir}-\xi^\mathrm{in}$). They also
find a second contribution called therein ``anharmonic
saturation term'' (see eq.\ (15) in ref.\ [20] or the third equation of
ref.\ [21]). Due to the $a^\dag
\bar{B}^\dag \bar{B}\bar{B}$ structure of the interaction from which it
appears, we could think it to be the photon-assisted exchange channel we
find. However, when considered carefully, the physics it bares is at odd.
This can be seen seen from the prefactor
$Y$ of the interaction term, eq.\ (15) of ref.\ [20], explicitly given in
eq.\ (16): It contains the
\emph{cube} of the exciton relative motion wave function. The three wave
functions barely come from the
three boson-exciton operators of $a^\dag \bar{B}^\dag \bar{B}\bar{B}$. 
Since the exciton-photon vacuum Rabi coupling also depends on the
exciton wave function, their exciton saturation
density
$n_{sat}$ as defined in ref.\ [21], ends by reading
\begin{equation}
\frac{1}{n_{sat}L^2}=\frac{2\sum_{\v k}\phi_0^3(\v k)}{\sum_{\v
k}\phi_0(\v k)}=\frac{4\pi}{7}\left(\frac{a_X}{L}\right)^2\ ,
\end{equation}
where $\phi_0(\v k)=\langle\v
k|\nu_0\rangle$ is the 2D ground
state exciton wave function equal to
$\sqrt{2\pi}[1+(ka_X/2)^2]^{-3/2}a_X/L$, with $a_X$ being the 3D Bohr
radius and $L^2$ the well area. 

On the opposite, in our photon-assisted
exchange scattering $\chi$, enters the exciton Pauli scattering.
For ground state exciton with zero center-of-mass momentum, it reads
[22] 
\begin{equation}
\lambda\left(^{\nu_0\v 0\ \nu_0\v 0}_{\nu_0\v 0\ \nu_0\v
0}\right)=\sum_{\v k}\phi_0^4(\v
k)=\frac{4\pi}{5}\left(\frac{a_X}{L}\right)^2\ .
\end{equation}
Therefore we see that the photon-assisted
exchange scattering $\chi$ has to definitely contain the
$4^{\mathrm{th}}$ power of the exciton wave function, due to the
\emph{four} excitons involved in the carrier exchange linked with this
channel. Since ref.\ [20] is very elliptical with respect to the
derivation of their interaction Hamiltonian, we cannot
point out the origin of the incorrectness.

\noindent \textbf{Microcavity polaritons}

We end this work on polaritons by considering a specific
example, microcavity polaritons, as they are of high current interest for
the possible observation of Bose-Einstein condensation of semiconductor
excitations [17,18].

In microcavities, the excitons are localized in quantum well. In the 
strong confinement regime, they are quasi-2D excitons; their energy
simply reads
$E_{\nu\v q}=(E_g+\varepsilon_{ze}+\varepsilon_{zh}+\epsilon_{\nu}+
\hbar^2q^2/2M_X)$, where $\v q$ is the 2D exciton center-of-mass momentum.
To get the polariton scatterings from the exciton scatterings, we first
need to determine the exciton-polariton and photon-polariton overlaps. To
do so, we note that, for a single photon mode
$n=\v q$  close to the ground state exciton resonance, we can forget the
higher quantum well subbands as well as the exciton excited levels, so
that the excitons of interest reduce to
$i=(\nu_0,\v q)$. Due to momentum conservation in the exciton-photon
coupling, the eigenvalue equation for the polariton $(P_{\v
q},\mathcal{E}_{\v q})$ then reduces to a
$2\times 2$ matrix
\begin{equation}
\left( \begin{array}{cc} E_{\nu_0\v q}-\mathcal{E}_{\v q} &
\Omega^\ast_{\v q}\\
                        \Omega_{\v q} & \omega_{\v
q}-\mathcal{E}_{\v q}
\end{array} \right)
\left( \begin{array}{c} \beta_{\nu_0\v q,P_{\v q}} \\ \alpha_{\v q,P_{\v
q}}
\end{array}
\right) =0\ ,
\end{equation}
where $\omega_{\v
q}=\sqrt{(k^2+q^2)c^2/\epsilon}$ is the cavity photon energy with
$k=\pi/L_{cav}$ for an optical cavity length
$L_{cav}$. The vacuum Rabi coupling between the cavity photon with energy
$\omega_{\v q}$ and the ground state exciton $(\nu_0\v q)$ located in the
well reduces to $\Omega_{\v q}=\Omega_0\sqrt{\omega_{\v 0}/\omega_{\v
q}}$. Typical values for InGaAs/GaAs microcavities are 
$\omega_{\v 0}\simeq E_{\nu_0\v 0}\simeq$ 1.5 eV, $\Omega_0\simeq 4$
meV, while
$m_e\simeq 0.067$, $m_h\simeq 0.34$ and $\epsilon\simeq 13$. This gives a
3D Bohr radius $a_X\simeq 12$ nm while $e^2/\epsilon a_X\simeq 8.8$ meV.
These parameters, used in the present paper, correspond to
Savvidis \emph{et al.}'s experimental conditions [23].

The resulting frequencies for the upper and lower polariton branches
are given by
\begin{equation}
\mathcal{E}_{\v q}^{\pm}=(\omega_{\v q}+E_{\nu_0\v q}\pm\sqrt{T_{\v
q}})/2
\ ,
\end{equation} 
where $T_{\v q}=(\omega_{\v q}-
E_{\nu_0\v q})^2+4|\Omega_{\v q}|^2$, the
corresponding overlaps being
\begin{eqnarray}
\beta_{\nu_0\v q,P_{\v q}}^- &=& \alpha_{P_{\v q},\v
q}^+=2\Omega^\ast_{\v q}/N_{\v q }\nonumber\\
\alpha_{\v q,P_{\v q}}^- &=& -\beta_{\nu_0\v q,P_{\v q}}^+=(\omega_{\v
q}-E_{\nu_0\v q}-\sqrt{T_{\v q}}) /N_{\v q}\ ,
\end{eqnarray} 
with $N_{\v q}^2=(\omega_{\v q}-E_{\nu_0\v q}-\sqrt{T_{\v q}})^2+
4|\Omega_{\v q}|^2$.

We use these results to calculate the Coulomb and
photon-assisted exchange scatterings for a pair of
identical cavity polaritons $(\v q,\v q)$ in the lower branch,
scattered into a signal polariton $(\v q=\v 0)$ at the
zone center and an idler polariton $(2\v q)$. Figure 2(a) shows the
dispersion relation $\mathcal{E}_{\v q}^{\pm}$ of the two polariton
branches, while fig.2(b) shows the energy difference $\Delta_{\v q}=
(2\mathcal{E}_{\v q}^- -\mathcal{E}_{\v 0}^- -\mathcal{E}_{2\v q}^-)$
associated to the $(\v q,\v q)\rightarrow (\v 0,2\v q)$ transition.
Energy conservation in the transition rate given in eq.\ (14) imposes
$q=q^\ast$ with $q^\ast\simeq 0.025/a_X$. This momentum inside the well
corresponds to work at the ``magic'' angle
$\theta=15^\circ$ for the laser beam outside the cavity.
As the photon momenta are small
on the $a_X^{-1}$ exciton scale, the momenta
$(\v q,\v q)$ and
$(\v 0,2\v q)$ involved in this
transition
are essentially zero on this scale.
The exciton Coulomb and Pauli
scatterings appearing in the polariton scatterings 
can thus be replaced  by their values for zero center-of-mass
momentum. Consequently, the $\v q$ dependence of these
cavity-polariton scatterings only comes from the
polariton overlaps. By noting that
$\xi^\mathrm{dir}\left(^{n\ j}_{i\ i}\right)=0$, as seen from eq.\ (B.18)
in ref.\ [9], --- which makes the direct Coulomb channel reducing to
zero, --- the ratio of the contributions to the polariton transition rate
from
$(\v q,\v q)$ to
$(\v 0,2\v q)$ coming from the photon-assisted exchange channel and from
the na\"{\i}ve Coulomb channel,then reads
\begin{eqnarray}
R_{\v q}&=&\frac{\chi\left(^{2\v q\ \v q}_{\,\v 0\ \,\,\v q}\right)+
\chi\left(_{2\v q\ \v q}^{\v 0\ \ \v q}\right)}{\tilde{\xi}^\mathrm{in}
\left(^{2\v q\ \v q}_{\,\v 0\ \,\,\v q}\right)}\nonumber\\
&\simeq&
\frac{\lambda\left(^{\nu_0\v 0\
\nu_0\v 0}_{\nu_0\v 0\ \nu_0\v 0}\right)}{\xi^\mathrm{in}
\left(^{\nu_0\v 0\
\nu_0\v 0}_{\nu_0\v 0\ \nu_0\v 0}\right)}
\left(\Omega_{\v 0}\frac{
\alpha_{P_{\v 0},\v 0}^-}{\beta_{P_{\v 0},\nu_0\v 0}^-}
+\Omega_{2\v q}\frac{\alpha_{P_{2\v q},2\v q}^-}
{\beta_{P_{2\v q},\nu_02\v q}^-}\right)\ ,
\end{eqnarray}
The
diagonal ``in'' Coulomb scattering for 2D ground state excitons with zero
center-of-mass momentum can be calculated analytically as
\begin{equation}
\xi^\mathrm{in}\left(^{\nu_0\v 0\
\nu_0\v 0}_{\nu_0\v 0\ \nu_0\v
0}\right)=-(4\pi-315\pi^3/1024)(a_X/L)^2(e^2/\epsilon a_X)\ ,
\end{equation}
\emph{i.e.}, $\xi^\mathrm{in} \simeq
-3.0(a_X/L)^2(e^2/\epsilon a_X)$.
By using the Pauli scattering given in eq.\ (16) and the overlaps
given in eq.\ (19), we obtain the ratio $R_{\v q}$ shown in fig.2(c).

It can be of interest to note that, for resonant photon $\omega_{\v 0}=
E_{\nu_0\v 0}$,
\begin{equation}
R_{q=0}=-2\Omega_0\,\frac{\lambda\left(^{\nu_0\v 0\
\nu_0\v 0}_{\nu_0\v 0\ \nu_0\v 0}\right)}{\xi^\mathrm{in}
\left(^{\nu_0\v 0\
\nu_0\v 0}_{\nu_0\v 0\ \nu_0\v 0}\right)}=2R_{q\rightarrow\infty}\ ,
\end{equation}
since, due to eq.\ (19), $\alpha_{P_{\v 0},\v 0}^-=-\beta_
{P_{\v 0},\nu_0\v 0}^-=-1/\sqrt{2}$, while $\alpha_{P_{2\v q},2\v q}^-$
goes to 0 when $q\rightarrow\infty$, assuming an infinite exciton
mass. 

We see from fig.2(c) that the photon-assisted exchange
channel produces a significant enhancement of the
polariton transition rate, when compared to the one coming from the
na\"{\i}ve Coulomb interactions between the excitonic components of the
polaritons: For typical experimental conditions as the ones considered
here, the increase of the transition rate at the ``magic'' value
$q=q^\ast$, 
\begin{eqnarray}
\frac{\left|\tilde{\xi}^\mathrm{in}\left(^{2\v q\ \v q}_{\v 0\ \ \v
q}\right)+\chi\left(^{2\v q\ \v q}_{\v 0\ \ \v q}\right)+\chi
\left(^{\v 0\ \ \v q}_{2\v q\ \v q}\right)\right|^2}
{\left|\tilde{\xi}^\mathrm{in}\left(^{2\v q\ \v q}_{\v 0\ \ \v
q}\right)\right|^2}\nonumber\\
=
\left(1+R_{\v q}\right)^2\simeq 2.1\ ,
\end{eqnarray}
is found to be slightly larger than 2.

\noindent\textbf{Conclusion}

Through a fully microscopic procedure, which makes use of the
composite-exciton many-body theory we have recently proposed, it is now
possible to approach the strong coupling of photons and
semiconductor excitations in an exact way, \emph{i.e.}, without mapping
the excitons into a boson subspace at any stage. We show that the
composite nature of the polaritons gives rise to a photon-assisted
exchange scattering, free from Coulomb process. The results of this exact
approach disagree with the ones obtained by using bosonized excitons,
even through the elaborate procedure proposed by the Quattropani's group.
This photon-assisted exchange channel, dominant when one of the
scattered polaritons has a strong photon character, produces a
significant enhancement of the transition rate for a pair of pump
microcavity polaritons scattered into an idler and a signal at the zone
center. This paper also contains crucial tools to tackle the many-body
physics of polaritons, towards their possible Bose Einstein condensation.

\newpage

\begin{figure}
\centering
\includegraphics[scale = 0.8]{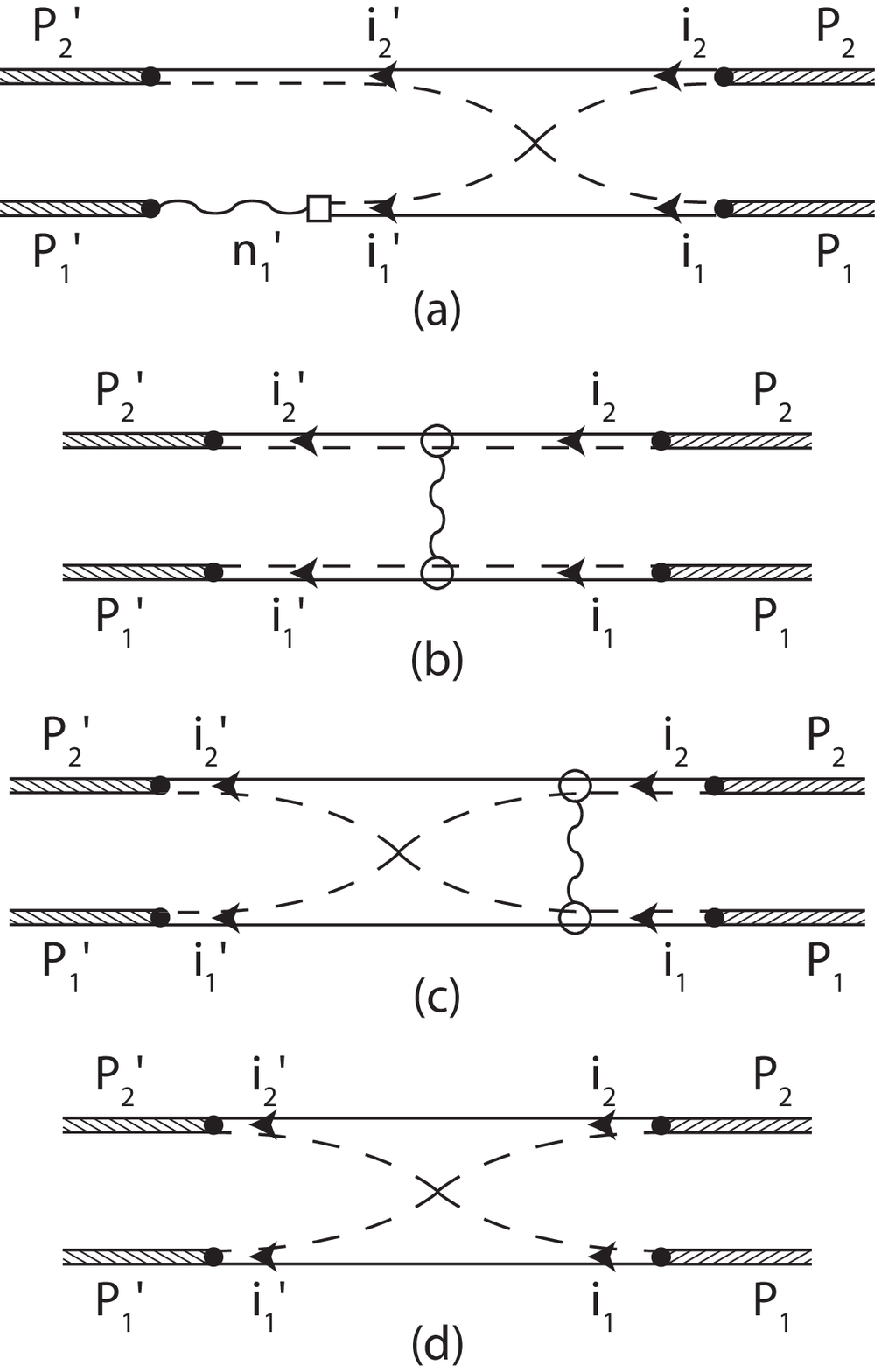}
\caption{(a): Photon-assisted exchange scattering of two polaritons 
$\chi\left(^{P'_2\ P_2}_{P'_1\ P_1}\right)$. (b):
Direct Coulomb scattering of two polaritons
$\tilde{\xi}^\mathrm{dir}\left(^{P'_2\ P_2}_{P'_1\ P_1}\right)$. (c):
``In'' exchange Coulomb scattering of two polaritons
$\tilde{\xi}^\mathrm{in}\left(^{P'_2\ P_2}_{P'_1\ P_1}\right)$. (d):
Pauli scattering of two polaritons $\tilde{\lambda}\left(^{P'_2\
P_2}_{P'_1\ P_1}\right)$.}
\end{figure}

\newpage

\begin{figure}
\centering
\includegraphics[scale = 0.75]{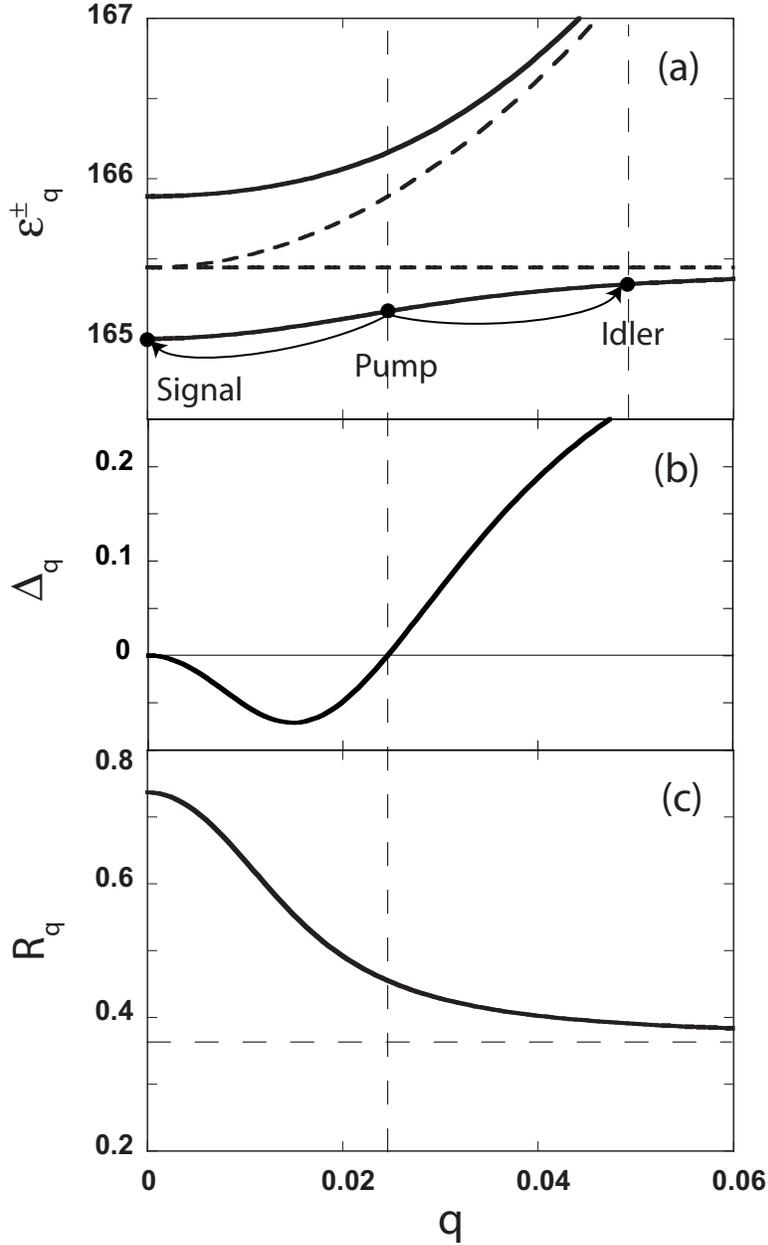}
\caption{(a) Dispersion $\mathcal{E}^{\pm}_{\v q}$ of the lower and upper
microcavity polariton branches, as given in eq.\ (24), for the typical
experimental conditions described in the text. The scattering of two pump
polaritons at the ``magic'' angle into a signal polariton at $\v q=\v 0$
and an idler at
$2\v q$ is also sketched. (b) Detuning
$\Delta_{\v q}$ associated to this scattering process, as a function of
the pump wave vector $\v q$. We see that energy is conserved for
$q=0.025$, which corresponds to a magic pump angle outside the
microcavity $\theta\simeq 15^{\circ}$. (c): Ratio $R_{\v q}$ of the photon
assisted exchange scattering and the Coulomb scattering, defined in eq.\
(26) for the transition rate of two polaritons
$(\v q,\v q)$ scattered into $(\v 0,2\v q)$. In the figure, the energies
are in $e^2/\epsilon a_X$ unit and the momenta in $a_X^{-1}$ unit.} 
\end{figure}


\begin{thebibliography}{99}
\bibitem{19}
To get eq.\ (14), we have used eq.\ (12) and the fact that 
$\sum_{P'_1,P'_2}\tilde{\lambda}\left(^{P''_2\ P'_2}
_{P''_1\ P'_1}\right)\,\chi\left(^{P'_2\ P_2}_{P'_1\ P_1}\right)$ is
equal to zero, which follows from $\langle v|B_i\alpha_n^\dag|v\rangle=0$,
the sums over $P$'s being performed through closure relations.
\end{thebibliography}
\end{document}